\documentclass[11pt,a4paper]{article}

\usepackage{amsfonts}

\usepackage{graphicx}
\usepackage{amsmath,amssymb}
\usepackage{epsfig}
\usepackage{dsfont}
\usepackage{color}
\usepackage{xcolor}
\usepackage{mathrsfs}
\usepackage{subfig}

\def\nbgroup{N}
\def\cardi{n_i}
\def\temps{T_{ij}}
\def\censure{C_{ij}}
\def\yobs{X_{ij}}
\def\bX{\mathbf{X}}

\def\dobs{\Delta_{ij}}

\def\hazard{h_{ij}(t|b_i)}

\def\baseline{h_0(t)}

\def\desfixe{Z_{ij}}
\def\desalea{W_{ij}}

\def\fixe{\beta}

\def\frailty{b_i}
\def\frailtyv{\mathbf{ b}}

\def\be{\xi}

\newcommand{\Rset}[1]{\mbox{$\mathbb{R}^{#1}$}}

\def\1{1\!{\mathrm l}}

\def\bX{\bold{X}}
\def\bdelta{\bold{\Delta}}

\newcommand{\norm}[1]{\left\lVert#1\right\rVert}

\newtheorem{theorem}{Theorem}
\newtheorem{remark}{Remark}
\newtheorem{lemma}[theorem]{Lemma}

\title{Convergent stochastic algorithm for parameter estimation in frailty models using integrated partial likelihood}

\author{Oodally Ajmal$^1$, Duchateau Luc$^2$, Estelle Kuhn$^1$
}
\date{ $^1$  MaIAGE, INRA, Universit\'e Paris-Saclay,  78350 Jouy-en-Josas, France.\\
$^2$Ghent University, Faculty of Veterinary Medicine Department of Nutrition, Genetics and Ethology, Belgium.
}

\begin{document}

\maketitle 







\begin{abstract}
Frailty models are often the model of choice for heterogeneous survival data. A frailty model contains both random effects and fixed effects, with the random effects accommodating for the correlation in the data. Different estimation procedures have been proposed for the fixed effects and the variances of and covariances between the random effects. Especially with an unspecified baseline hazard, i.e., the Cox model, the few available methods  deal only with a specific correlation structure. In this paper, an estimation procedure, based on the integrated partial likelihood, is introduced, which can generally deal with any kind of correlation structure. The new approach, namely the maximisation of the integrated partial likelihood, combined with a stochastic estimation procedure allows also for a wide choice of distributions for the random effects. First, we demonstrate the almost sure convergence of the stochastic algorithm towards a critical point of the integrated partial likelihood. Second, numerical convergence properties are evaluated by simulation. Third, the advantage of using an unspecified baseline hazard is demonstrated through application on cancer clinical trial data.
\end{abstract}

%
%
%


\section{Introduction}
\label{S:1}

Survival analysis consists in the analysis of the time of occurence of an event of interest. The Cox model introduced in \cite{Cox72} is often used in this area. It allows us to model the risk of occurence of the event of interest, also called hazard,  as the product of a baseline hazard function and a function of the covariates. The regression coefficients are usually estimated by maximisation of the partial likelihood which does not depend on the baseline hazard function. The good asymptotic properties of the estimator, namely the consistency,  asymptotic normality and efficiency, based on partial likelihood are detailed and proved in \cite{Andersen82}. However, an underlying assumption, called proportional hazard assumption, of this model is that the ratio of the hazards of two individuals is constant over time. This assumption is quite strong and often not fulfilled in practice due to a lack of homogeneity in real data. For example, in a clinical study, the data may be clustered into groups based on the location of the clinics or on different medical staves involved in collecting samples. Frailty models introduced by Vaupel et al in \cite{Vaupel79} allow to do away with this assumption by taking into account heterogeneity through non observed random effects. For more details on frailty models, we refer to \cite{DuchateauBook08}, \cite{wienke2010frailty}.


The literature on parameter estimation in frailty models is quite rich. Maximum likelihood estimation based on an Expected Maximization (EM) algorithm has been studied by \cite{andersen92} with frailties following a Gamma distribution in both non and semi-parametric models. The asymptotic properties of the maximum likelihood estimates with a plug-in estimator for the baseline hazard from a Gamma frailty model without covariates have been studied by \cite{Murphy94}, \cite{Murphy95} and for a correlated Gamma frailty distribution by \cite{Parner98}. The choice of the Gamma distribution is here motivated by its mathematical convenience. Indeed, a closed form of the marginal likelihood can be calculated when the frailties are assumed to follow a Gamma distribution.

An approach based on the maximization of a penalized partial likelihood with a Laplace approximation of the marginal likelihood has been proposed by \cite{Ripatti00}. Also, \cite{Duchateau04} implemented an approach based on penalized partial likelihood using an iterative algorithm based on the marginal and penalized log likelihoods. A semi-parametric approach where the baseline hazard is estimated with a splines basis in the Gamma frailty model is implemented in the R package \textit{frailtypack} developed by \cite{frailtypack}. An estimation method based on the first and second order Laplace approximations of the complete partial likelihood has been proposed by \cite{korea17} and implemented in the R package \textit{frailtyHL}. However to the best of our knowledge, none of these existing algorithms has been proved to be convergent theoretically.

The aim of this paper is to propose an efficient  stochastic algorithm to maximize the integrated partial likelihood  and to prove its theoretical convergence property. We consider the criteria given by   the integrated partial likelihood for the frailty model and the  estimator that maximizes this criteria. We present an efficient stochastic EM algorithm to calculate its value. Then we establish its theoretical almost sure convergence  to a critical point of the integrated partial likelihood. Moreover, we highlight the benefit of using the integrated partial likelihood through simulation studies and real data analysis. 

The paper is organized as follows. Section 2 deals with the frailty model. The integrated partial likelihood and the estimator associated are presented in Section 3. The algorithmic method for inference is presented in Section 4. An extended frailty model and the corresponding stochastic estimation procedure whose convergence property is established are detailed in Section 5. The simulation and real data studies are presented in Section 6. The paper ends with a conclusion and a  discussion. 

\section{The Frailty Model}
\label{sec:model}
\subsection{Description of the model}

We consider a population of individuals clustered into $\nbgroup$ groups. We denote by $\cardi$ the size of the $i$-th group for $1 \leq i \leq \nbgroup$. We denote the event time and censoring time of the individual $j$ in group $i$ by $\temps$ and $\censure$ respectively for $1 \leq i \leq \nbgroup$ and $1 \leq j \leq \cardi$. We observe the variable $\yobs=\min(\temps,\censure)$ and the censoring indicator  defined as $\dobs=\mathds{1}_{\{\temps \leq \censure\}}$. We denote by $\bX=(\yobs)_{1 \leq i \leq \nbgroup, 1 \leq j\leq \cardi}$ and by $\bdelta=(\dobs)_{1 \leq i \leq \nbgroup, 1 \leq j\leq \cardi}$  the observations.

We consider the following frailty model where the hazard for the individual $j$ of group $i$ is expressed as follows : 

\begin{equation}
\label{eq:model}
\forall t \geq 0 \ \hazard = \baseline \ \exp (\desfixe^t \fixe + \desalea^t \frailty),  
\end{equation}
 where  $h_0(t)$ denotes the baseline hazard function at time $t$, 
  $\desfixe$ and $\desalea$  the covariates of individual $j$ of group $i$,  $\fixe \in \Rset{b}$  the unknown regression parameter vector and $b_i$ $\in \Rset{f}$  the common frailty      shared by individuals of group $i$. We assume that the probability density function of the unobserved frailty is parametric and denote by  $\gamma$ its parameter taking values in $\Rset{c}$.

Therefore  the model parameters are $h_0$, $\fixe$ and $\gamma$.
The parameter of interest is usually $\fixe$, enabling the quantification of the effects of the covariates which is often the main objective of real data analysis. 

\subsection{Assumptions on the model}
We introduce  the following usual assumptions on the frailty model: \\
\textbf{(F1)} The  censoring times $(\censure)$ are independent of the event times $(\temps)$ and of the frailties $(\frailty)$. \\
\textbf{(F2)} Conditionally to the frailties $(\frailty)$, the event times $(\temps)$ are independent. \\
\textbf{(F3)} The frailties $(\frailty)$ are identically and independently distributed having common density $g_\gamma$. \\
\textbf{(F4)} The function $h_0$ belongs to the set of functions defined on $\mathbb{R}^{+}$ taking values in $\mathbb{R}^{+}$. \\
\textbf{(F5)} The probability density function  of the frailties denoted by $g_{\gamma}$ belongs to the set of curved exponential family of probability density functions where $\gamma$ takes values in $\mathbb{R}^c$. 

\begin{remark}
We note here that \textbf{(F4)} is required only for the construction of the partial likelihood. The regularity condition is weaker than the one in \cite{kuhn13} where a choice of parametric structure is made on the baseline hazard function.
\end{remark}


\section{Estimation by maximisation of the integrated partial likelihood}
We consider the criteria  defined by the integrated partial likelihood  for the frailty model following the idea of \cite{Cox72}. Then we define the estimator as the parameter value that maximises this criteria. 

\subsection{Definition of the integrated partial likelihood}
Following the idea used in the Cox model, we consider the conditional partial likelihood defined  as follows:
\begin{equation}
\label{eq:likepartcond}
L_{\text{cond}}^p(\theta;\bX,\bdelta \mid \frailtyv) = \prod_{i=1}^{N} \prod_{j=1}^{n_i} \Bigg( \frac{\exp(\desfixe^t \beta +\desalea^t b_i)}{\sum_{(l,k) \in R(X_{(ij)})} \exp(Z_{lk}^t \beta + W_{lk}^t b_l)} \Bigg)^{\Delta_{ij}}
\end{equation}
where $\theta = (\beta, \gamma)$ $\in \mathbb{R}^b \times \mathbb{R}^c$, $R(X_{(ij)}) = \left\lbrace 1 \leq l \leq N, 1 \leq k \leq n_l : X_{lk} \geq X_{(ij)} \right\rbrace$ is the set of individuals still at risk at time $X_{(ij)}$ and $\frailtyv=(b_i)_{1 \leq i \leq N}$.  \\

We then easily deduce the complete partial likelihood expression:
\begin{eqnarray}
\label{eq:likepartfull}
&L^p(\theta;\bX,\bdelta,\frailtyv) = \prod_{i=1}^{N} g_{\gamma} (b_i)\\ 
&\times \prod_{i=1}^{N} \prod_{j=1}^{n_i} \Bigg( \frac{\exp(\desfixe^t \beta +\desalea^t b_i)}{\sum_{(l,k) \in R(X_{(ij)})} \exp(Z_{lk}^t \beta + W_{lk}^t b_l)} \Bigg)^{\Delta_{ij}} \nonumber
\end{eqnarray}
We emphasize that this partial likelihood no longer involves the baseline $h_0$ as the partial likelihood in the Cox model.

Finally we define the integrated partial likelihood as defined in \cite{therneau18}, also called marginal partial likelihood,  obtained by integrating the complete partial likelihood over the unobserved frailties $\frailtyv$:
\begin{equation}
\label{eq:obslike}
L_{\text{marg}}^p (\theta;\bX,\bdelta) = \int L^p(\theta;\bX,\bdelta,\frailtyv) d \frailtyv
\end{equation}

\begin{remark}
We recall that as in the Cox model, this integrated partial likelihood is not a likelihood function, but  acts as one as explained in \cite{Ripatti00}.
\end{remark}

\subsection{Definition of the maximum integrated partial likelihood estimate}
\label{sec:esti}
Following \cite{therneau18}, we define the estimator $\hat{\theta}$ for the parameters vector as the value that maximizes the integrated partial likelihood:
\begin{equation}
\label{eq:miple}
\hat{\theta} = \underset{\theta}{\text{argmax}} \ L_{\text{marg}}^p (\theta;\bX,\bdelta)
\end{equation}


If there exists an analytical expression of the integrated partial likelihood, it  can  be maximized directly.  When the computation of the integrated partial likelihood is difficult, an EM type algorithm can be implemented for the maximization procedure. Therefore, we propose to calculate $\hat{\theta}$ by using  a stochastic Expectation Maximization (EM) algorithm following in the footsteps of \cite{kuhn13}.
 
 \section{Algorithmic methods for inference}

\subsection{Description of the algorithm  for parameter estimation}

We consider the stochastic EM algorithm introduced by \cite{KuhnLavielle04} to evaluate the estimator of the parameters defined in Equation (\ref{eq:miple}). It is an extension of the stochastic approximation EM algorithm developed by \cite{dely99} where the
EM algorithm is coupled with a Markov Chain Monte Carlo (MCMC) procedure to simulate the unobserved frailties. 

Each iteration of the algorithm is composed of three  steps  detailed below. We start with initial values $\theta_0$,$\frailtyv_0$ and $Q_0$ arbitrarily chosen.

Repeat until convergence for $k \geq 1$:

\begin{enumerate}
\item \textbf{Simulation step}: draw realizations $\frailtyv_k$  of the unobserved frailties 
 \begin{equation*}
      \frailtyv_k   \sim \Pi_{\theta_{k-1}} (\mathbf{b}_{k-1},.)
    \end{equation*}
   where    $\Pi_{\theta_{k-1}}$ is  a  transition probability of a convergent Markov chain  having as stationary distribution  the conditional distribution  $\pi^p_{\theta_{k-1}} (. \vert \bX,\bdelta)$  defined by
$$\pi^p_{\theta} (\frailtyv \vert \bX,\bdelta) = \frac{L^p(\theta;\bX,\bdelta,\frailtyv)}{\int L^p(\theta;\bX,\bdelta,\frailtyv) d \textbf{b}}$$
\item \textbf{Stochastic approximation step}: compute for all $\theta$
\begin{equation}
\label{eq:Qeqn}
Q_k (\theta) = Q_{k-1} (\theta) + \mu_k \left(\log L^p(\theta;\bX,\bdelta,\frailtyv_k) - Q_{k-1} (\theta) \right) 
\end{equation}
where the sequence $(\mu_k)$ satisfies
\begin{equation*}
0 \leq \mu_k \leq 1 \text{,            } \sum \mu_k = +\infty \text{,            } \sum \mu_k^2 < +\infty
\end{equation*}
\item \textbf{Maximisation step}:
update the parameter estimate
$$\theta_{k} = \underset{\theta}{\text{argmax }} Q_k(\theta)$$
\end{enumerate}

This algorithm will be called Algorithm 1 below. Further practical details on the algorithm, namely the simulation procedure used to sample  the realizations of the unobserved frailties $\frailtyv_k$, the computation of the quantity $Q_k$ and the update of parameter estimates $\theta_{k}$ can be found in Appendices A and B. The choice of the stepsize sequence  $(\mu_k)_k$ is detailed in Section 6.

\subsection{Estimation of the Fisher Information Matrix}
We consider the usual estimate of the Fisher Information Matrix, namely the observed Fisher information matrix $I_{obs}(\theta) = - \partial_{\theta}^2 \text{log} L_{\text{marg}}^p (\theta;\bX,\bdelta) $ (see \cite{andersen97}). Using Louis's missing information principle (see \cite{louis}), we express the matrix $I_{obs}(\theta)$ as:

\begin{align*}
I_{obs}(\theta) &= - \mathbb{E}_{\theta} \big( \partial_{\theta}^2 \text{log} L^p (\theta;\bX,\bdelta, \textbf{b}) \mid \bX,\bdelta \big) - \\
&\text{Cov}_{\theta} \big( \partial_{\theta} \text{log} L^p (\theta;\bX,\bdelta, \textbf{b}) \mid \bX,\bdelta \big)
\end{align*}
where  $\mathbb{E}_{\theta}$ and $\text{Cov}_{\theta}$ denote respectively the expectation and the covariance under the posterior distribution $\pi_{\theta}^p$ of the frailty.

We approximate the quantity $I_{obs}(\theta)$ by a Monte Carlo sum based on the realizations of the Markov chain generated in the algorithm having as stationary distribution the posterior distribution $\pi_{\theta}^p$. After a  burn-in period, we use the remaining M realizations $(\textbf{b}_m)_{1 \leq m \leq M}$ of the Markov chain to compute the following quantity:

\begin{align*}
\hat{I}_M(\hat{\theta}) &= - \frac{1}{M} \sum_{m=1}^M  \partial_{\theta}^2 \text{log} L^p (\hat{\theta};\bX,\bdelta, \textbf{b}_m) \\
& - \frac{1}{M} \sum_{m=1}^M \big( \partial_{\theta} \text{log} L^p (\hat{\theta};\bX,\bdelta,\textbf{b}_m) \times \partial_{\theta} \text{log} L^p (\hat{\theta};\bX,\bdelta,\textbf{b}_m)^t \big) \\
& + \frac{1}{M^2} \bigg( \sum_{m=1}^M \partial_{\theta} \text{log} L^p (\hat{\theta};\bX,\bdelta,\textbf{b}_m) \bigg) \times \bigg( \sum_{m=1}^M \partial_{\theta} \text{log} L^p (\hat{\theta};\bX,\bdelta,\textbf{b}_m) \bigg)^t
\end{align*}

The ergodic theorem in \cite{meyn} guarantees the convergence of the quantity $\hat{I}_M(\hat{\theta})$ to the observed Fisher information matrix $I_{obs}(\hat{\theta})$ as M goes to infinity. 

\section{Extended frailty model and convergent estimation algorithm}

Most of the theoretical convergence properties of stochastic 
 EM like algorithms have been established in the case of curved exponential families as for examples in \cite{dely99, KuhnLavielle04, stephkuhn}. Since the complete partial likelihood defined in (\ref{eq:likepartfull}) does not belong to the curved exponential family of probability density functions, we introduce an extended frailty model. 

\subsection{Extended frailty model}
We consider an extended frailty model where  the regression parameter $\beta$ is considered as a population random variable. The extended latent variables are denoted by $\xi = (b_i, i=1, \dots, n, \beta)$. 
Moreover we assume that the population variable $\beta$ follows a Gaussian distribution with unknown expectation $\beta_0$ and fixed variance $\sigma_{\beta}^2$. We  denote the new parameters to be estimated by $\eta = (\beta_0, \gamma)$.
The complete likelihood corresponding to the model can be written as follows:

\begin{align}
    \label{eq:elikelihood}
    \begin{split}
    &L^e(\eta;\bX,\bdelta,\be) = \prod_{i=1}^{N} g_{\gamma} (b_i) f_{\beta_0} (\beta) \\ 
&\prod_{i=1}^{N} \prod_{j=1}^{n_i} \Bigg( \frac{\exp(\desfixe^t \beta +\desalea^t b_i)}{\sum_{(l,k) \in R(X_{(ij)})} \exp(Z_{lk}^t \beta + W_{lk}^t b_l)} \Bigg)^{\Delta_{ij}}
\end{split}
\end{align}
where $f$ stands for the Gaussian probability density function.
 This likelihood function belongs to the curved exponential family as soon as the frailty probability density function $g_{\gamma}$  belongs to the curved exponential family.  Sufficient statistics are explicit and can be expressed as $S(\xi) = \Big( \sum_{i=1}^N S_{f} (b_i), \beta \Big)$ where $S_f(b_i)$ are  sufficient statistics corresponding to the frailties $(b_i)$.

%
%





By assumption \textbf{(F5)}, the complete partial  likelihood defined in (\ref{eq:elikelihood}) can  be written as follows:
\begin{equation*}
 L^e(\eta;\bX,\bdelta,\be) =  \text{exp} (-\Psi(\eta) + \langle S (\be), \Phi(\eta) \rangle )
\end{equation*}
where $S$, $\Psi$ and $\Phi$ are  Borel functions .

%

\subsection{Description of the stochastic EM Algorithm with truncation on random boundaries}

Following \cite{stephkuhn}, we detail a seoncd algorithm, called Algorithm 2, based on the extended likelihood. 

Let $(\mathcal{K}_{q})_{q} \geq 0$ be a sequence of increasing compact subsets of $S$ such as $\bigcup_{q \geq 0} \mathcal{K}_{q} = S$ and $\mathcal{K}_q \subset \text{int}(\mathcal{K}_{q+1})$, $\forall q \geq 0$. \\


Initialize $\eta_0$ in $\Theta$, $\be_0$ and $s_0$ in two fixed compact sets $\textbf{K}$ and $\mathcal{K}_0$ respectively. \\

Repeat until convergence for $k \geq 1$:
\begin{enumerate}
    \item \textbf{Simulation step:} Draw $\bar{\be}$ from a  kernel $\Pi_{\eta_{k-1}}$ of a convergent Markov chain having as stationary distribution the conditional distribution with the current parameters:
    \begin{equation*}
        \bar{\be} \sim \Pi_{\eta_{k-1}} (\be_{k-1},.)
    \end{equation*}
    \item \textbf{Stochastic approximation step:} Compute 
    \begin{equation*}
        \bar{s} = s_{k-1} + \mu_{k} (S(\bar{\be}) - s_{k-1})
    \end{equation*}
    \item \textbf{Truncation step:} If $\bar{s}$ is outside the current compact set $\mathcal{K}_{\kappa_{k-1}}$, where $\kappa$ is the index of the current active truncation set, or too far from the previous value $s_{k-1}$ then restart the stochastic approximation in the initial compact set, extend the truncation boundary to $\mathcal{K}_{\kappa_{k}}$ and start again with a bounded value of the missing variable. Otherwise, set $(\be_k,s_k) = (\bar{\be},\bar{s})$ and keep the truncation boundary to $\mathcal{K}_{\kappa_{k-1}}$. 
    \item \textbf{Maximization step:} 
    \begin{equation*}
        \eta_k = \underset{\eta}{\text{argmax}} \lbrace - \Psi (\eta) + \langle s_k, \Phi (\eta) \rangle \rbrace
    \end{equation*}
\end{enumerate}



In this second algorithm, we construct a sequence $(\be_k,s_k)$ while satisfying two conditions at each iteration $k$. Namely we check whether the stochastic approximation wanders outside the current compact set and whether the current value is not too far from the previous value. The latter can be expressed as follows:
\begin{equation*}
    \norm{s_k - s_{k-1}} \leq \epsilon_{k}
\end{equation*}
where $\epsilon = (\epsilon_k)_{k \geq 0}$ is a monotone non-increasing sequence of positive numbers. A more detailed description of the truncation step can be found in \cite{moulines05}.

\subsection{Convergence property of Algorithm 2 in the extended frailty model}


We consider classical assumptions required to prove the convergence of EM like algorithms as following those of \cite{dely99}.  \\

\textbf{(M3)} The function $\Bar{s}: \Theta \rightarrow S$ defined as:

\begin{equation*}
    \Bar{s} (\eta) = \int_{\mathbb{R}^{l}} |S (\be)| \pi_s^e (\be) d \be
\end{equation*}

where 

\begin{equation*}
    \pi_s^e (\be) = \frac{\text{exp} (- \Psi (\eta(s)) + \langle s, \Phi (\eta(s)) \rangle )}{\int \text{exp} (- \Psi (\eta(s)) + \langle s, \Phi (\eta(s)) \rangle ) d\be}
\end{equation*}

 is continuously differentiable on $\Theta$.

\textbf{(M4)} The function $l^e: \Theta \rightarrow \mathbb{R}$ defined as the marginal extended log-likelihood

\begin{align*}
    l^e(\eta) = \text{log} \int_{\mathbb{R}^{l}} L^e(\eta;\bX,\bdelta,\be) d \be
\end{align*}
is continuously differentiable on $\Theta$ and 
\begin{equation*}
    \partial_{\eta} \int_{\mathbb{R}^{l}} L^e(\eta;\bX,\bdelta,\be) d \be = \int_{\mathbb{R}^{l}} \partial_{\eta} \  L^e(\eta;\bX,\bdelta,\be) d \be
\end{equation*}

\textbf{(M5)} There exists a function $\hat{\eta}: S \rightarrow \Theta$ s.t:
$$\forall \ s \in S, \forall \ \eta \in \Theta, \ L(\hat{\eta}(s);s) \geq L(\eta;s)$$

where $L: S \times \Theta \rightarrow \mathbb{R}$ is defined as

\begin{equation}
\label{eq:L_s}
    L(\eta;s) = \text{exp} (- \Psi (\eta) + \langle s, \Phi (\eta) \rangle )
\end{equation}

Moreover, the function $\hat{\eta}$ is continuously differentiable on S. \\

Following in the lines of \cite{moulines05}, we state a first assumption \textbf{(A1')} that guarantees the existence of a global Lyapunov function denoted by $w$ defined as:

\begin{equation}
\label{eq:w_s}
    w(s) = - \text{log} \int L^e (\hat{\eta} (s); \bX,\be) d\be
\end{equation} 

for the mean field $h$ defined as: 

\begin{equation}
\label{eq:h_s}
    h(s) = \int (S(\be) - s) \pi_s^e (\be) d\be
\end{equation}

\textbf{(A1')} The functions $w$ and $h$ are such that
\begin{enumerate}
    \item[(i)] there exists an $M_0 > 0$ such that 
    \begin{equation*}
        \mathcal{S} \overset{\Delta}{=} \lbrace s \in S, \langle \nabla w(s),h(s)\rangle = 0 \rbrace \subset \lbrace s \in S, w(s) < M_0 \rbrace
    \end{equation*}
    where $w$ is defined in (\ref{eq:w_s}) and $h$ is defined in (\ref{eq:h_s}).
    \item[(ii)] there exists $M_1 \in ]M_0,\infty]$ such that $\lbrace s \in S, w(s) < M_1 \rbrace$ is a compact set.
    \item[(iii)] the closure of $w(\mathcal{L})$ has an empty interior.
\end{enumerate}

\textbf{(A4)} The sequences $\mu = (\mu_k)_{k \geq 0}$ and $\epsilon = (\epsilon_k)_{k \geq 0}$ are non-increasing, positive and satisfy $\sum_{k=0}^{\infty} \mu_k = \infty$, $\underset{k \rightarrow \infty}{\text{inf}} \epsilon_k = 0$ and $\sum_{k=1}^{\infty} \lbrace \mu_k^2 + \mu_k \epsilon_k^a + (\mu_k \epsilon_k^{-1})^p \rbrace < \infty$, where $a \in ]0,1]$ and $p \geq 2$. \\

Finally we consider the usual drift assumption \textbf{(DRI)} which are detailed in  \cite{moulines05}. 

\begin{theorem}
\label{mainThm}
Assume that \textbf{(F1-F5)}, \textbf{(M3-M5)}, \textbf{(A1')}, \textbf{(A4)} and \textbf{(DRI)} are fulfilled. Then we have with probability 1 
\begin{equation*}
    \underset{k \rightarrow \infty}{\text{lim}} d(\eta_k, \mathcal{L}) = 0
\end{equation*}
where $(\eta_k)_k$ is generated by Algorithm \textbf{2}, $d(x,A)$ denotes the distance from $x$ to any closed subset $A$  and $\mathcal{L} = \lbrace \eta \in \Theta, \partial_{\eta} \text{ log } L_{\text{marg}}^e (\eta; \bX, \bdelta) = 0 \rbrace$.
\end{theorem}

The assumption \textbf{(A1')} corresponds to the assumptions \textbf{(A1)} (i), (ii), (iv) of \cite{moulines05} respectively. Assumption \textbf{(A4)} deals with the conditions on the step-size sequences involved in  the stochastic approximation and truncation steps of Algorithm 2.\\ 

\paragraph{\textbf{Proof of Theorem \ref{mainThm}}: }
We will first apply Theorem $5.5$ of of \cite{moulines05} to prove the convergence of the sequence $(s_k)$ and checked therefore the assumptions required. To prove that assumption (\textbf{A1)(iii)} of \cite{moulines05} is fulfilled  in our case, we
 establish the following lemma following the lines of the proof of Lemma 2 of \cite{dely99} using in our case the partial likelihood instead of the likelihood:
\begin{lemma}
\label{lemma:A13}
 Assuming \textbf{(F1–F5)} and \textbf{(M3-M5)}, we have for any $s \in S \setminus \mathcal{S} \ \langle \nabla w(s),h(s)\rangle < 0$~
\end{lemma}

\paragraph{\textbf{Proof of Lemma 2}} 
Assumption \textbf{(M1)}  of \cite{dely99} is implied by \textbf{(F5)}. 
 To fulfill assumption \textbf{(M2)} of \cite{dely99}, it suffices to show that $\Psi$ and $\Phi$ are twice continuously differentiable. This is a straight consequence of assumptions \textbf{(F1–F5)}. The end of the proof follows the same lines as Lemma 2 of \cite{dely99}. \\

Thereby  assumption (\textbf{A1)(iii)} of \cite{moulines05} is fulfilled  in our case. 
 As detailed in \cite{moulines05},   assumptions  \textbf{(DRI)}  imply assumptions \textbf{(A2-A3)}  by Proposition 6.1. Thus we can 
apply Theorem $5.5$ of \cite{moulines05}. We get   that the sequence $(s_k)$ 
generated by Algorithm $2$ satisfayes $\lim_k d(s_k, \mathcal{S})=0$. Following the lines of the proof of Lemma 2 of \cite{dely99}, we get that $\lim_k d(\eta_k, \mathcal{L})=0$. The proof of Theorem \ref{mainThm} is therefore complete.\\

\section{Numerical studies}
All numerical studies have been done using R version 3.3.1 on an Intel Core i7-8550U CPU @ 1.99 GHz, 16 GB RAM.

The aim of our numerical experiments is to compare the performances of the Maximum Integrated Partial Likelihood (MIPL) estimator defined in Section \ref{sec:esti}  to those of other  estimators existing in the literature. We also  analyse a real  dataset of bladder cancer. \\

We run both algorithms in the numerical studies. Since we get results of the same order, we only present the ones obtained using Algorithm 1,  the main motivation of the extended model and of Algorithm 2 being the   theoretical convergence result. 

\begin{enumerate}
    \item The decreasing positive step size $(\mu_k)$ is taken as follows: \\
\begin{eqnarray*}
 \forall k > K_0, & \  \mu_k &= \frac{1}{(k-K_0)}  \\
    \forall k \leq  K_0,& \ \mu_k &= 1\\
\end{eqnarray*}    
   
   where  $K_0$ is a number to be specified. The algorithm is said to have no memory during the first  $K_0$ iterations. After this burn-in time which allows for the algorithm to visit the parameter space, the sequence $(\mu_k)_k$ decreases and converges to zero as $k \rightarrow \infty$. 
    \item The transition kernel used for simulating the unobserved frailty is chosen as a transition kernel of a Metropolis Hastings algorithm with  proposal distribution $q$ equal to  a Gaussian distribution centered at the current value $\mathbf{b_{k-1}}$ at the $k^{th}$ iteration.
    \item We define a stopping criterion based on the relative difference between the values of the parameters for two consecutive iterations. Let us fixed a positive  threshold $\epsilon > 0$. If for some $k>1$:
    \begin{equation*}
        \frac{\norm{\theta_k - \theta_{k-1}}}{\norm{\theta_{k-1}}} < \epsilon
    \end{equation*}
    holds true for for three  consecutive iterations, the algorithm is stopped. We set $\epsilon = 10^{-4}$ in the simulation study.
\end{enumerate}


\subsection{Simulated data}

We consider the following setting. The frailties $(b_i)$  are drawn from a centered normal distribution with variance $\gamma = 0.7$. The regression parameter $\beta$ used to simulate the data is chosen equal to the vector $(2,3)$ of size $2$. The covariates $((\desalea,\desfixe))$ are generated independently according to a Bernoulli distribution.   We consider  varying number $N$ of clusters and $n_i=4$ observations per cluster.

\subsubsection{Study of the consistency property of $\hat{\theta}$}
We begin by studying numerically the consistency of the estimate $\hat{\theta}$. The Weibull baseline hazard defined as $h_0(t)=\lambda  \rho t^{\rho-1}$  for $t>0$ is considered in this section using the parameter values $\lambda=0.01$ and $\rho=1.5$. There is no censoring.

\begin{equation}
\label{weibullmodel}
\forall t \geq 0 \  \hazard = \lambda \rho t^{\rho-1} \ \exp (\desfixe^t \fixe +  \frailty), \hspace{3mm} \lambda > 0, \rho > 0
\end{equation}

The estimate $\hat{\theta}$ is evaluated using the algorithm described in Section 4.1. 

\begin{table}[h]
\begin{center}
\label{table1}
\caption{Mean of parameter estimates  $\hat{\theta}$ and standard deviation in parenthesis  obtained from 500 repetitions with different group sizes. The number of observations per group is fixed at $n_i = 4$.}
\begin{tabular}[H]{|l|l|l|l|l|}
\hline
Parameters & True values &  N=10 & N=20 & N=50 \\
\hline
\hline
$\beta_1$ & 2 & 1.794 & 1.996 & 2.002 \\
& & (0.385) & (0.353) & (0.320) \\
\hline
$\beta_2$ & 3 & 2.652 & 2.995 & 2.999 \\
& & (0.427) & (0.390) & (0.339) \\
\hline
$\gamma$ & 0.7 & 0.490 & 0.649 & 0.707 \\
& & (0.656) & (0.477) & (0.287) \\
\hline
\end{tabular}
\end{center}
\end{table}

The results supporting the numerical consistency of $\hat{\theta}$ are displayed in Table 1. $N$ refers to the number of groups. As the the number of groups $N$ progressively increases, the corresponding estimates  get closer to the true values and the standard deviation decreases. 

\subsubsection{Comparing the maximum integrated partial likelihood estimate with a parametric estimate}
We consider a parametric estimate defined  in the model  with a Weibull baseline hazard function defined as $h_0(t) = \lambda \rho t^{\rho -1}, \ \lambda > 0, \rho > 0$. We denote the vector of parameters by $\theta_{\text{weibull}} = (\lambda,\rho,\fixe,\gamma)$. The expression of the complete likelihood is given by:

\begin{equation}
\begin{split}
& L^{\text{weibull}}(\theta_{\text{weibull}};\bX,\bdelta,\frailtyv) = \prod_{i=1}^{N} g_{\gamma} (b_i) \prod_{i=1}^{N} \prod_{j=1}^{n_i} \big( \lambda \rho \yobs^{\rho - 1} \\ & \times \exp(\desfixe^t \beta + b_i) \big)^{\Delta_{ij}} \exp (-\lambda \yobs^{\rho} \exp(\desfixe^t \beta + b_i))
\end{split}
\end{equation}

The marginal likelihood is obtained by integrating over the frailties $b$ : 

\begin{equation}
L_{\text{marg}}^{\text{weibull}}(\theta;\bX,\bdelta) = \int L^{\text{weibull}}(\theta_{\text{weibull}};\bX,\bdelta,\frailtyv) d\frailtyv
\end{equation}

We denote by $\hat{\theta}_{\text{weibull}}$ the estimator of the maximum of the  marginal likelihood : 

\begin{equation}
\hat{\theta}_{\text{weibull}} = \underset{\theta_{\text{weibull}}}{\text{argmax}} \  L_{\text{marg}}^{\text{weibull}}(\theta_{\text{weibull}};X,\Delta)
\end{equation}

The value of $\hat{\theta}_{\text{weibull}}$ is computed using the MCMC-SAEM algorithm proposed in \cite{kuhn13}. The event times are first simulated according to ($\ref{weibullmodel}$) with Weibull parameters $\lambda=0.01$ and $\rho=1.5$. The number of groups N is fixed at a value of 250. There is no censoring. The results are presented in Table 2. We conclude that both methods give good estimates in this example. 

We then consider event times simulated from the model  using a Gompertz baseline hazard function. The modeling equation is as follows:

\begin{equation}
\label{gompertzmodel}
\hazard =  \lambda \exp (\alpha t) \ \exp (\desfixe^t \fixe +  \frailty), \hspace{3mm} \lambda > 0, \alpha > 0
\end{equation}

The event times are simulated according to ($\ref{gompertzmodel}$) with Gompertz parameters $\lambda=0.08$ and $\alpha=2$. There is no censoring. The results are presented in Table 3. The estimate $\hat{\theta}$ which does not require any modeling assumption of $h_0$ proves to be a good estimator where as $\hat{\theta}_{\text{weibull}}$ does not give good results as it can be seen in Table 3. The wrong specification of $h_0$ for the latter introduces bias in the estimation of the parameters. 
 These results therefore show the advantages of not having to model the baseline hazard $h_0$ in the estimation procedure. 

\begin{table}
\begin{center}
\label{table2}
\caption{Mean of parameter estimates and model-based standard error in parentheses obtained from 500 repetitions with the event times following a Weibull distribution to compare the parametric estimate to the partial integrated likelihood estimate. $N = 250$ and $n_i = 4$.}
\begin{tabular}[H]{|l|l|l|l|l|}
\hline
Method & $\beta_1$ &  $\beta_2$ & $\gamma$ \\
\hline
\hline
True values & 2 & 3 & 0.7 \\
\hline
$\hat{\theta}$ & 2.033 & 3.056 & 0.702 \\
 & (0.133) & (0.121) & (0.106) \\
\hline
$\hat{\theta}_{\text{weibull}}$ & 1.982 & 2.944 & 0.701 \\
 & (0.133) & (0.145) & (0.111) \\
\hline
\end{tabular}
\end{center}
\end{table}

\begin{table}
\begin{center}
\label{table3}
\caption{Mean of parameter estimates and  model-based standard error in parentheses  obtained from 500 repetitions with the event times following a Gompertz distribution to compare the parametric estimate to the integrated partial likelihood estimate. $N = 250$ and $n_i = 4$.}
\begin{tabular}[H]{|l|l|l|l|}
\hline
Method & $\beta_1$ &  $\beta_2$ & $\gamma$ \\
\hline
\hline
True values & 2 & 3 & 0.7 \\
\hline
$\hat{\theta}$ & 2.031 & 3.041 & 0.732 \\
 & (0.126) & (0.134) & (0.129) \\
\hline
$\hat{\theta}_{\text{weibull}}$ & 1.380 & 2.029 & 0.270 \\
 & (0.112) & (0.146) & (0.126) \\
\hline
\end{tabular}
\end{center}
\end{table}


\subsubsection{Comparison of the maximum integrated partial likelihood estimate with other estimates}
The aim of the following simulation study is to compare the performances of the maximum integrated partial likelihood estimate with those of other estimates. Therefore we consider  the estimate based on penalized partial likelihood implemented in the R package \textit{coxme} based on \cite{Ripatti00} and two estimates based on the h-likelihood implemented in the R package \textit{frailtyHL} detailed in \cite{korea17}. \\

The estimation in the \textit{coxme} package is based on the maximisation of a penalized partial likelihood. This estimator is denoted by $\hat{\theta}_{\text{coxme}}$ later. \\

The h-likelihood methods implemented in \textit{frailtyHL} are based on a Laplace approximation of the marginal partial likelihood which is then maximised. The two estimators based on h-likelihood chosen differ in the order of the Laplace approximations. They are denoted by $\hat{\theta}_{\text{HL(0,1)}}$ and $\hat{\theta}_{\text{HL(1,2)}}$ with the first one based on the first order Laplace approximation and the second one based on the second order Laplace approximation. \\

\subsubsection{Effect of censoring level on parameter estimation}
We first investigate the effect of censoring when comparing the different estimation procedures. We recall that there was no censoring in the previous simulation settings. The event times were simulated according to ($\ref{weibullmodel}$) with Weibull parameters $\lambda=0.01$ and $\rho=1.5$. The number of groups N is fixed at a value of 250. Data are simulated under two different censoring levels, low (Table 4) and moderate (Table 5). \\

In Table 4, in the low censoring level case, we observe that the MIPL estimate $\hat{\theta}$ and the estimate $\hat{\theta}_{\text{HL(1,2)}}$ seem to be closer to the true values as opposed to the estimates $\hat{\theta}_{\text{coxme}}$ and $\hat{\theta}_{\text{HL(0,1)}}$. We make the same observation in Table 5 with the  moderate censoring. It seems that $\hat{\theta}$ and $\hat{\theta}_{\text{HL(1,2)}}$ have the same performance level in the estimation of $\beta$ and give better estimates than $\hat{\theta}_{\text{coxme}}$ and $\hat{\theta}_{\text{HL(0,1)}}$. We note however that $\hat{\theta}$ gives slightly better estimates than $\hat{\theta}_{\text{HL(1,2)}}$ for the variance $\gamma$ for both low and moderately censored settings.

\begin{table}
\begin{center}
\label{table4}
\caption{Mean of parameter estimates and model-based standard error in parentheses  obtained from 500 repetitions with the event times following a Weibull distribution and a low censoring level of 20 $\%$. Comparsion of MIPL with \textit{coxme} and \textit{frailtyHL}. $N = 250$ and $n_i = 4$.}
\begin{tabular}[H]{|l|l|l|l|}
\hline
Method & $\beta_1$ &  $\beta_2$ & $\gamma$ \\
\hline
\hline
True Values & 2 & 3 & 0.7 \\
\hline
$\hat{\theta}$ & 1.968 & 2.968 & 0.672 \\
 & (0.123) & (0.156) & (0.116) \\
\hline
$\hat{\theta}_{\text{coxme}}$ & 1.922 & 2.901 & 0.606 \\
 & (0.120) & (0.151) & (0.107) \\
\hline
$\hat{\theta}_{\text{HL(0,1)}}$ & 1.930 & 2.939 & 0.607 \\
 & (0.118) & (0.155) & (0.107) \\
\hline
$\hat{\theta}_{\text{HL(1,2)}}$ & 1.954 & 2.976 & 0.647 \\
 & (0.120) & (0.158) & (0.117) \\
\hline
\end{tabular}
\end{center}
\end{table}

\begin{table}
\begin{center}
\label{table5}
\caption{Mean of parameter estimates and model-based standard error in parentheses  obtained from 500 repetitions with the event times following a Weibull distribution and a moderate censoring level of 40 $\%$. Comparsion of MIPL with \textit{coxme} and \textit{frailtyHL}. $N = 250$ and $n_i = 4$.}
\begin{tabular}[H]{|l|l|l|l|}
\hline
Method & $\beta_1$ &  $\beta_2$ & $\gamma$ \\
\hline
\hline
True Values & 2 & 3 & 0.7 \\
\hline
$\hat{\theta}$ & 1.896 & 2.859 & 0.641 \\
 & (0.133) & (0.153) & (0.120) \\
\hline
$\hat{\theta}_{\text{coxme}}$ & 1.850 & 2.791 & 0.575 \\
 & (0.125) & (0.149) & (0.084) \\
\hline
$\hat{\theta}_{\text{HL(0,1)}}$ & 1.847 & 2.808 & 0.576 \\
 & (0.125) & (0.150) & (0.113) \\
\hline
$\hat{\theta}_{\text{HL(1,2)}}$ & 1.873 & 2.846 & 0.615 \\
 & (0.126) & (0.151) & (0.121) \\
\hline
\end{tabular}
\end{center}
\end{table}

\subsubsection{Robustness to misspecification of the frailty distribution}
We investigate in this section the case where the frailty distribution is misspecified in the estimating procedure. For example, assuming a normal frailty as  done previously  when the frailties instead follow a mixture of normal distributions might introduce bias in the estimates. We study the effects of a misspecification of the frailty distribution on the four estimators presented above. We first consider data  simulated with a multiplicative Gamma frailty term.  We observe that  all  estimating procedures give good estimations when a normal frailty is assumed for the estimation task (results non presented). Then we consider frailties  drawn from a mixture of normal distributions as follows:

\begin{equation*}
    b \sim \frac{1}{2}\mathcal{N}(-10,2) + \frac{1}{2}\mathcal{N}(10,2)
\end{equation*}

In all estimating procedures, a normal frailty is assumed. The event times were simulated according to ($\ref{weibullmodel}$) with Weibull parameters $\lambda=0.01$ and $\rho=1.5$. The number of groups N is fixed at a value of 250 and there are 4 observations per cluster. All event times are non-censored. The results are presented in Table 6. We observe that the estimates obtained with our method denoted by $\hat{\theta}$ and with \textit{frailtyHL} are close to the true value where as the one obtained by \textit{coxme} does not adjust well to the misspecification of the frailty distribution leading to some bias in the estimation of $\beta$ in this example. \\

\begin{table}
\begin{center}
\label{table6}
\caption{Mean of parameter estimates and model-based standard error in parentheses  obtained from 500 repetitions with the event times following a Weibull distribution.  Comparison of MIPL estimate with \textit{coxme} and \textit{frailtyHL} estimates when the frailty distribution is misspecified. A mixture of Gaussian  frailties is used to simulate the dataset whereas a Gaussian frailty is assumed in the estimation procedure. $N = 250$ and $n_i = 4$.}
\begin{tabular}[H]{|l|l|l|l|l|}
\hline
Method & $\beta_1$ &  $\beta_2$ & $\gamma$ \\
\hline
\hline
$\hat{\theta}$ & 2.036 & 3.040 & 25.5 \\
& (0.163) & (0.201) & (0.743) \\
\hline
$\hat{\theta}_{\text{coxme}}$ & 1.531 & 2.304 & 6.079 \\
& (0.124) & (0.133) & (0.566) \\
\hline
$\hat{\theta}_{\text{HL(1,2)}}$ & 2.022 & 3.019 & 23.0 \\
& (0.110) & (0.128) & (2.96) \\
\hline
\end{tabular}
\end{center}
\end{table}

\subsubsection{Correlated frailties}
In all of the previous simulation studies, the shared frailty model with the frailty acting only on the group level has been considered. In order to apply the MIPL estimating procedure on a real cancer dataset detailed in Section 6.2, we consider the modeling of the hazard function as follows:

\begin{equation}
\label{corfrailty}
    \hazard = \baseline \ \exp (b_{0i} + \desfixe^t (\fixe + b_{1i}))
\end{equation}

with $b = (b_{0},b_{1}) \sim \mathcal{N} (0, \Sigma)$ where $\Sigma$ = $\begin{pmatrix} \sigma_0^2 & \sigma_{01} \\ \sigma_{01} & \sigma_1^2 \end{pmatrix}$ \\

We estimate the parameters $\theta = (\beta, \sigma_0^2, \sigma_1^2, \sigma_{01})$ by maximising the integrated partial likelihood. The event times are simulated following \eqref{corfrailty} with a Weibull baseline hazard parametrized by $\lambda = 0.01$, $\rho =1.5$ and the regression paramater $\beta = (2,3)$. The frailty variances $\sigma_0^2$ and $\sigma_1^2$ are set to 0.8 and 0.4 respectively and the covariance term $\sigma_{01}$ is set to 0.226. The number of observations per group is not the same for all groups. The group sizes are fixed so as to be close to the configuration of the groups in the real dataset. The results are presented in Table 7. We observe that the estimate $\hat{\theta}$ is close to the true values for all the parameters. The model standard errors based on the estimation of the observed Fisher information matrix are also very small. This conclusive simulation study allows for the estimating procedure to be  applied for analysing the real dataset. 

\begin{table}
\begin{center}
\label{table7}
\caption{Mean of parameter estimates and standard deviation of estimates obtained from 500 repetitions using Weibull baseline with $\lambda=0.01$ and $\rho=1.5$ in the corrrelated frailty model.}
\begin{tabular}[H]{|l|l|l|l|l|l|}
\hline
Method & $\beta_1$ &  $\beta_2$ & $\sigma_0^2$ & $\sigma_1^2$ & $\sigma_{01}$\\
\hline
\hline
True Values & 2 & 3 & 0.8 & 0.4 & 0.226 \\
\hline
$\hat{\theta}$ & 2.020 & 3.016 & 0.805 & 0.403 & 0.218 \\
 & (0.009) & (0.010) & (0.008) & (0.006) & (0.00004) \\
\hline
\end{tabular}
\end{center}
\end{table}

\subsection{Real data analysis}
We consider a bladder cancer dataset from the EORTC. A combined analysis was carried out of individual patient data from 2596 superficial bladder cancer patients included in seven European Organization for Research and Treatment of Cancer trials 30781, 30782, 30791, 30831, 30832, 30845, and 30863 (Genito-Urinary tract cancer Group). Only the groups with more than 20 patients were included in the dataset to be analyzed. After data processing, we are left with 39 groups of patients of different sizes. The censoring level is about 51 $\%$ and about 80 $\%$ of the patients follow an intravesical treatment (see \cite{sylvester}) which is the only covariate considered. The studies conducted on this dataset suggests that the treatment effect $b_{1i}$ might be correlated to the center effect $b_{0i}$. This leads us to model the hazard function as detailed in (\ref{corfrailty}).
 We estimate the parameters $\theta = (\beta, \sigma_0^2, \sigma_1^2, \sigma_{01})$ by maximising the integrated partial likelihood. We run the algorithm using a grid of initial values and the mean of the obtained estimates is computed. The results are then compared with the estimates, which we denote by $\hat{\theta}_{cst}$, obtained in \cite{kuhn13} where a constant baseline hazard is assumed.

\begin{figure}[htbp]
\centering
\subfloat[$\beta$ trajectories]{\label{fig:a}\includegraphics[width=0.45\linewidth]{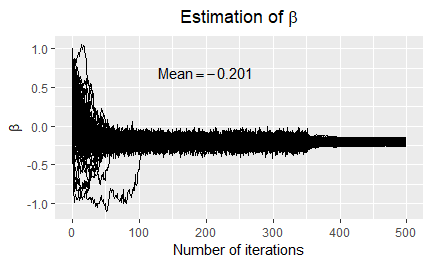}}\qquad
\subfloat[$\sigma_0^2$ trajectories]{\label{fig:b}\includegraphics[width=0.45\linewidth]{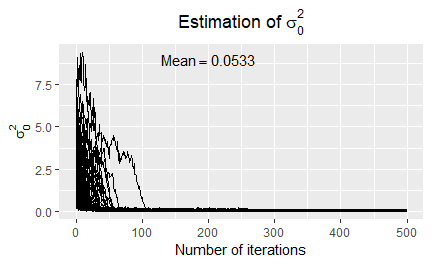}}\\
\subfloat[$\sigma_1^2$ trajectories]{\label{fig:c}\includegraphics[width=0.45\linewidth]{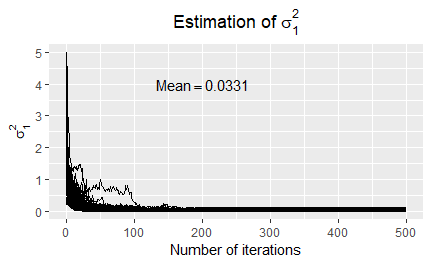}}\qquad
\subfloat[$\sigma_{01}$ trajectories]{\label{fig:d}\includegraphics[width=0.45\linewidth]{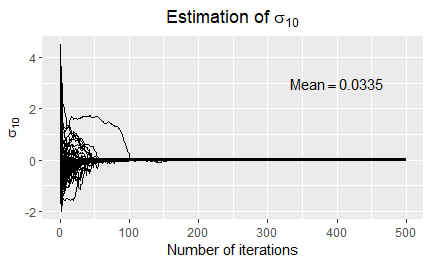}}%
\caption{Integrated partial likelihood estimates for EORTC bladder cancer dataset.}
\label{fig:eortc_t}
\end{figure}

The trajectories of all parameters estimated are shown in Figure \ref{fig:eortc_t}. We observe that whatever the initial conditions, all trajectories seem to lead to more or less the same values. The algorithm is therefore not sensible to initial conditions. The estimates obtained in \cite{kuhn13} are presented in the second column of Table 7. The estimates obtained with $\hat{\theta}$ and $\hat{\theta}_{cst}$ are not close, especially the parameter of interest $\beta$ and the variance $\sigma_1^2$ which takes into account the effect of the treatment. Thus, in this example, choosing a parametric constant form for the baseline affects strongly the estimation of the parameter of interest $\beta$, leading to possible wrong interpretation of the effect of the covariates. On the other hand, the estimate $\hat{\theta}$ does not rely on  any parametric assumption on the baseline, leading to a robust estimation procedure with respect to any parametric choice for the baseline.

\begin{table}
\label{fig:eortc}
\begin{center}
\label{table8}
\caption{Mean of the estimates and mean model-based standard error in parentheses obtained with $\hat{\theta}_p$ and $\hat{\theta}_{cst}$ on the EORTC bladder cancer dataset.}
\begin{tabular}[H]{|l|l|l|}
\hline
Parameters & $\hat{\theta}_p$ & $\hat{\theta}_{cst}$ \\
\hline
\hline
$\beta$ & -0.206 & -0.254 \\
 & (0.007) & (0.070) \\
\hline
$\sigma_0^2$ & 0.0712 & 0.0306 \\
 & (0.0001) & (0.0002) \\
\hline
$\sigma_1^2$ & 0.0435 & 0.107 \\
 & (0.0002) & (0.0006) \\
\hline
$\sigma_{01}$ & 0.0428 & 0.0573 \\
 & (0.000001) & (0.000003) \\
\hline
\end{tabular}
\end{center}
\end{table}

\section{Conclusion and discussion}
We consider as estimation criteria the integrated partial likelihood and the corresponding estimate which maximizes this criteria. The main advantage of this criteria is that it does not depend on any choice of the baseline function. We propose a  stochastic approximation EM algorithm coupled with a MCMC procedure for calculating the parameter estimates in practice. The almost sure convergence of this algorithm to a critical point of the integrated partial likelihood is established under classical assumptions. We then validate the performance of the estimation procedure through simulation studies which highlight the good  properties for finite sample size. In cases where the  baseline hazard function is misspecified, the proposed  estimate performs better than the parametric one. When the hazard function is correctly specified, we perform just as good. The proposed estimate called MIPL is then compared to existing estimates in the literature namely given by the \textit{coxme} and \textit{frailtyHL} packages where the intractable integral is approximated through a Laplace approximation. The robustness of all estimates to a misspecification of the frailty distribution is analyzed. The simulation setting also takes into account light to heavy censoring to see how the different estimates perform. Finally, we analyse a real bladder cancer dataset and compare our results with a parametric estimating procedure from the literature. \\

Since we have proposed an efficient convergent algorithm to compute the MIPL estimate, it would be now of great interest to study its  asymptotic properties as consistency, asymptotic normality and efficiency.

\appendix

\section{Description of the simulation procedure used to sample realizations  for the unobserved frailties }
We usually construct $\Pi_{\theta}$ as a step of a Metropolis Hastings algorithm with proposal distribution  $q$. Sample a candidate $b^c$: \\
$$b^c \sim q(. \vert b_{k-1}; \theta_{k-1})$$ \\
We then calculate the acceptance ratio : 
\begin{equation*}
\alpha(b_{k-1},b^c) = \text{min} \bigg( 1,\frac{\pi_{\theta_{k-1}}(b^c \vert \bX,\bdelta) q(b_{k-1} \vert b^c ; \theta_{k-1})}{\pi_{\theta_{k-1}}(b_{k-1} \vert \bX,\bdelta) q(b^c \vert b_{k-1} ; \theta_{k-1})} \bigg)
\end{equation*}
The simulated candidate is accepted with probability $\alpha(b_{k-1},b^c)$.
$$
b_{k} =
\begin{cases}
b^c & \text{with probability $\alpha(b_{k-1},b^c)$}\\
b_{k-1} & \text{otherwise}
\end{cases}
$$

\section{Estimation equations for the MIPL estimate}
We can rewrite (\ref{eq:Qeqn}) for an easier computation of the derivatives. By induction on $k$, we obtain:
\begin{align*}
Q_k (\theta) = Q_0 \prod_{i=1}^{k} (1 - \mu_i) + \sum_{i=1}^{k} \bigg( \mu_i \log L^p(\theta;\bX,\bdelta,b_i) \\ \times \prod_{\shortstack{$\scriptstyle j=i+1 $\\$\scriptstyle i<k$}}^{k} (1 - \mu_j) \bigg)
\end{align*}
This expression of $Q_k (\theta)$ makes the computation of the derivative with respect to $\theta$ relatively straightforward :
\begin{align*}
\frac{\partial Q_k (\theta)}{\partial \theta} = \sum_{i=1}^{k} \bigg( \mu_i \frac{\partial \log L^p(\theta; \bX,\bdelta,b^i)}{\partial \theta} \prod_{\shortstack{$\scriptstyle j=i+1 $\\$\scriptstyle i<k$}}^{k} (1 - \mu_j) \bigg)
\end{align*}

The log complete partial likelihood required to compute the quantity $Q$ can be expressed as follows: 

\begin{align}
\label{eq:loglikepartfull}
\begin{split}
\log L^p(\theta;\bX,\bdelta,\frailtyv) = \sum_{i=1}^{N} \sum_{j=1}^{n_i} \Delta_{ij} \Bigg( \desfixe^t \beta + \desalea^t b_i - \\ \log \bigg( \sum_{(l,k) \in R(X_{(ij)})} \exp(Z_{lk}^t \beta + W_{lk}^t b_l) \bigg) \Bigg) + \sum_{i=1}^{N} \log(g_{\gamma} (b_i))
\end{split}
\end{align}

Differentiating ($\ref{eq:loglikepartfull}$) with respect to $\beta$, we obtain the following equations :
\begin{align*}
\frac{\partial \log L^p(\theta;\bX,\bdelta,\frailtyv)}{\partial \beta} = \sum_{i=1}^{N} \sum_{j=1}^{n_i} \Delta_{ij} \Bigg( \desfixe - \\ \frac{\sum_{(l,k) \in R(X_{(ij)})} Z_{lk} \exp(Z_{lk}^t \beta +  W_{lk}^t b_l)}{\sum_{(l,k) \in R(X_{(ij)})} \exp(Z_{lk}^t \beta + W_{lk}^t b_l)} \Bigg)
\end{align*}

The parameter $\gamma$ is easily updated as it is found only in the last term of (\ref{eq:likepartfull}). In many cases, we can update by direct computation the parameter estimate of $\gamma$ and use a classic gradient descent algorithm to update $\beta$.

\bibliographystyle{plain}
\bibliography{example1.bib}

\begin{thebibliography}{10}

\bibitem{stephkuhn}
S.~{Allassonni\`ere}, E.~{Kuhn}, and A.~{Trouv\'e}.
\newblock Construction of bayesian deformable models via a stochastic
  approximation algorithm: A convergence study.
\newblock {\em Bernoulli}, 16:641--678, 2010.

\bibitem{andersen97}
P.~{Andersen}, J.~{Klein}, K.~{Knudsen}, and R.~{Tabanera y Palacios}.
\newblock Estimation of variance in cox's regression model with shared gamma
  frailties.
\newblock {\em Biometrics}, 53:1475--1484, 1997.

\bibitem{Andersen82}
P.K. {Andersen} and R.D. {Gill}.
\newblock Cox's regression model for counting processes : a large sample study.
\newblock {\em Annals of Statistics}, 10:1100--1120, 1982.

\bibitem{moulines05}
C.~{Andrieu}, E.~{Moulines}, and P.~{Priouret}.
\newblock Stability of stochastic approximation under verifiable conditions.
\newblock {\em SIAM J. Control Optim}, 44:283--312, 2005.

\bibitem{Cox72}
D.R. {Cox}.
\newblock Regression models and life-tables.
\newblock {\em Journal of the Royal Statistical Society}, 34:187–--220, 1972.

\bibitem{dely99}
B.~Delyon, M.~Lavielle, and E.~Moulines.
\newblock Convergence of a stochastic approximation version of the {E}{M}
  algorithm.
\newblock {\em Ann. Statist.}, 27(1):94--128, 1999.

\bibitem{Duchateau04}
L.~{Duchateau} and P.~{Janssen}.
\newblock Penalized partial likelihood for frailties and smoothing splines in
  time to first insemination models for dairy cows.
\newblock {\em Biometrics}, 60:608--614, 2004.

\bibitem{DuchateauBook08}
L.~Duchateau and P.~Janssen.
\newblock {\em The Frailty Model}.
\newblock Springer-Verlag, New York, 2008.

\bibitem{korea17}
Il~Do Ha, John-Hyeon Jeong, and Youngjo Lee.
\newblock {\em Statistical Modelling of Survival Data with Random Effects}.
\newblock Springer, Singapore, 2017.

\bibitem{kuhn13}
E.~{Kuhn} and C.~{El-Nouty}.
\newblock On a convergent stochastic estimation algorithm for frailty models.
\newblock {\em Statistics and Computing}, 23:413--423, 2013.

\bibitem{KuhnLavielle04}
E.~{Kuhn} and M.~{Lavielle}.
\newblock Coupling a stochastic approximation version of em with an mcmc
  procedure.
\newblock {\em ESAIM: Probability and Statistics}, 8:115--131, 2004.

\bibitem{louis}
T.A. {Louis}.
\newblock Finding the observed information matrix when using the em algorithm.
\newblock {\em J. Roy. Statist. Soc. Ser. B}, 44:226--233, 1982.

\bibitem{meyn}
S.~P. {Meyn} and R.L. {Tweedie}.
\newblock Markov chains and stochastic stability. communications and control
  engineering series.
\newblock {\em Springer-Verlag London Ltd}, 1993.

\bibitem{Murphy94}
S.~A. {Murphy}.
\newblock Consistency in a proportional hazards model incorporating a random
  effect.
\newblock {\em Annals of Statistics}, 22:712--731, 1994.

\bibitem{Murphy95}
S.~A. {Murphy}.
\newblock Asymptotic theory for the frailty model.
\newblock {\em Annals of Statistics}, 23:182--198, 1995.

\bibitem{andersen92}
G.~G. {Nielsen}, R.~D. {Gill}, P.~K. {Andersen}, and T.~I.~A. {Sorensen}.
\newblock A counting process approach to maximum likelihood estimation in
  frailty models.
\newblock {\em Scand. J. Statist.}, 19:25--44, 1992.

\bibitem{Parner98}
E.~{Parner}.
\newblock Asymptotic theory for the correlated gamma-frailty model.
\newblock {\em Annals of Statistics}, 26:183--214, 1998.

\bibitem{Ripatti00}
S.~{Ripatti} and J.~{Palmgren}.
\newblock Estimation of multivariate frailty models using penalized partial
  likelihood.
\newblock {\em Biometrics}, 56:1016–--1022, 2000.

\bibitem{frailtypack}
V.~{Rondeau}, Y.~{Mazroui}, and J.~{Gonzalez}.
\newblock frailtypack: An r package for the analysis of correlated survival
  data with frailty models using penalized likelihood estimation or
  parametrical estimation.
\newblock {\em Journal of Statistical Software}, 47:1--28, 2012.

\bibitem{sylvester}
R.~J. Sylvester, A.~P. van~der Meijden, W.~Oosterlinck, J.~A. Witjes,
  C.~Bouffioux, L.~Denis, D.~W. Newling, and K.~Kurth.
\newblock {{P}redicting recurrence and progression in individual patients with
  stage {T}a {T}1 bladder cancer using {E}{O}{R}{T}{C} risk tables: a combined
  analysis of 2596 patients from seven {E}{O}{R}{T}{C} trials}.
\newblock {\em Eur. Urol.}, 49(3):466--465, Mar 2006.

\bibitem{therneau18}
T~Therneau.
\newblock Coxme and the laplace approximation.
\newblock {\em Technical report}, 2018.

\bibitem{Vaupel79}
J.~{Vaupel}, K.~{Manton}, and E.~{Stallard}.
\newblock The impact of heterogeneity in individual frailty on the dynamics of
  mortality.
\newblock {\em Demography}, 16:439–--454, 1979.

\bibitem{wienke2010frailty}
A.~Wienke.
\newblock {\em Frailty Models in Survival Analysis}.
\newblock Chapman \& Hall/CRC Biostatistics Series. CRC Press, 2010.

\end{thebibliography}







\end{document}